\documentclass[sigconf]{acmart}
\AtBeginDocument{%
  \providecommand\BibTeX{{%
    \normalfont B\kern-0.5em{\scshape i\kern-0.25em b}\kern-0.8em\TeX}}}

\setcopyright{none}
\renewcommand\footnotetextcopyrightpermission[1]{}
\usepackage{multirow, subcaption, mathtools}
\begin{document}

\title{Erasing Labor with Labor: Dark Patterns and Lockstep Behaviors on Google Play}

\author{Ashwin Singh}
\affiliation{%
  \institution{IIIT Hyderabad, India}
  \country{}}
\orcid{0000-0002-2596-9649}
\email{ashwin19.iiith@gmail.com}
  
\author{Arvindh Arun}
\affiliation{%
  \institution{IIIT Hyderabad, India}
  \country{}}
\orcid{0000-0003-3469-6539}
\email{arvindh.a@research.iiit.ac.in}
  
\author{Pulak Malhotra}
\affiliation{%
  \institution{IIIT Hyderabad, India}
  \country{}}
\orcid{0000-0002-2492-7981}
\email{pulak.malhotra@students.iiit.ac.in}

\author{Pooja Desur}
\affiliation{%
  \institution{IIIT Hyderabad, India}
  \country{}}
\email{pooja.desur@students.iiit.ac.in}

\author{Ayushi Jain}
\affiliation{%
  \institution{IIIT Delhi, India}
  \country{}}
\email{ayushi19031@iiitd.ac.in}

\author{Dueng Horng Chau}
\affiliation{%
  \institution{Georgia Institute of Technology, USA}
  \country{}}
\orcid{0000-0001-9824-3323}
\email{polo@gatech.edu}

\author{Ponnurangam Kumaraguru}
\affiliation{%
  \institution{IIIT Hyderabad, India}
  \country{}}
\email{pk.guru@iiit.ac.in}

\begin{abstract}
Google Play's policy forbids the use of incentivized installs, ratings, and reviews to manipulate the placement of apps. However, there still exist apps that incentivize installs for other apps on the platform. To understand how install-incentivizing apps affect users, we examine their ecosystem through a socio-technical lens and perform a mixed-methods analysis of their reviews and permissions. Our dataset contains 319K reviews collected daily over five months from 60 such apps that cumulatively account for over 160.5M installs. We perform qualitative analysis of reviews to reveal various types of dark patterns that developers incorporate in install-incentivizing apps, highlighting their normative concerns at both user and platform levels. Permissions requested by these apps validate our discovery of dark patterns, with over 92\% apps accessing sensitive user information. We find evidence of fraudulent reviews on install-incentivizing apps, following which we model them as an edge stream in a dynamic bipartite graph of apps and reviewers. Our proposed reconfiguration of a state-of-the-art microcluster anomaly detection algorithm yields promising preliminary results in detecting this fraud. We discover highly significant lockstep behaviors exhibited by reviews that aim to boost the overall rating of an install-incentivizing app. Upon evaluating the 50 most suspicious clusters of boosting reviews detected by the algorithm, we find (i) near-identical pairs of reviews across 94\% (47 clusters), and (ii) over 35\% (1,687 of 4,717 reviews) present in the same form near-identical pairs within their cluster. Finally, we conclude with a discussion on how fraud is intertwined with labor and poses a threat to the trust and transparency of Google Play.
\end{abstract}

%   <concept>
%       <concept_id>10002978.10003029</concept_id>
%       <concept_desc>Security and privacy~Human and societal aspects of security and privacy</concept_desc>
%       <concept_significance>300</concept_significance>
%       </concept>
%   <concept>
%       <concept_id>10002978.10002997</concept_id>
%       <concept_desc>Security and privacy~Intrusion/anomaly detection and malware mitigation</concept_desc>
%       <concept_significance>300</concept_significance>
%       </concept>
%   <concept>
%       <concept_id>10003456.10003462</concept_id>
%       <concept_desc>Social and professional topics~Computing / technology policy</concept_desc>
%       <concept_significance>100</concept_significance>
%       </concept>
%  </ccs2012>

% \ccsdesc[300]{Security and privacy~Human and societal aspects of security and privacy}
% \ccsdesc[300]{Security and privacy~Intrusion/anomaly detection and malware mitigation}
% \ccsdesc[100]{Social and professional topics~Computing / technology policy}


\maketitle

\section{Introduction}

Google Play lists over 2.89 million apps on its platform~\cite{stat1}. In the last year alone, these apps collectively accounted for over 111 billion installs by users worldwide ~\cite{stat2}. Given the magnitude of this scale, there is tremendous competition amongst developers to boost the visibility of their apps. As a result, developers spend considerable budgets on advertising, with expenditure reaching 96.4 billion USD on app installs in 2021~\cite{stat3}. Owing to this competitiveness, certain developers resort to inflating the reviews, ratings, and installs of their apps. The legitimacy of these means is determined by Google Play’s policy, under which the use of incentivized installs is strictly forbidden~\cite{google}. Some apps violate this policy by offering users incentive in the form of gift cards, coupons, and other monetary rewards in return for installing other apps; we refer to these as \textit{install-incentivizing apps}. Past work~\cite{farooqi2020} found that apps promoted on install-incentivizing apps are twice as likely to appear in the top charts and at least six times more likely to witness an increase in their install counts. While their work focuses on measuring the impact of incentivized installs on Google Play, our work aims to develop an understanding of how it affects the \textit{users} of install-incentivizing apps. To this end, we perform a mixed-methods analysis of the reviews and permissions of install-incentivizing apps. Our ongoing work makes the following contributions:
\begin{enumerate}
    \item We provide a detailed overview of various dark patterns present in install-incentivizing apps and highlight several normative concerns that disrupt the welfare of users on Google Play. 
    \item We examine different types of permissions requested by install-incentivizing apps to discover similarities with dark patterns, with 95\% apps requesting permissions that access restricted data or perform restricted actions
    \item We show promising preliminary results in algorithmic detection of fraud and lockstep behaviors in reviews that boost overall rating of install-incentivizing apps, detecting near-identical review pairs in 94\% of the 50 most suspicious review clusters.
    \item We release our dataset comprising 319K reviews written by 301K reviewers over a period of five months and 1,825 most relevant reviews with corresponding qualitative codes across 60 install-incentivizing apps.~\cite{Ashwin}
\end{enumerate}

\begin{figure}[!h]
\centering
\begin{minipage}{.48\textwidth}
  \centering
    \includegraphics[width=\textwidth]{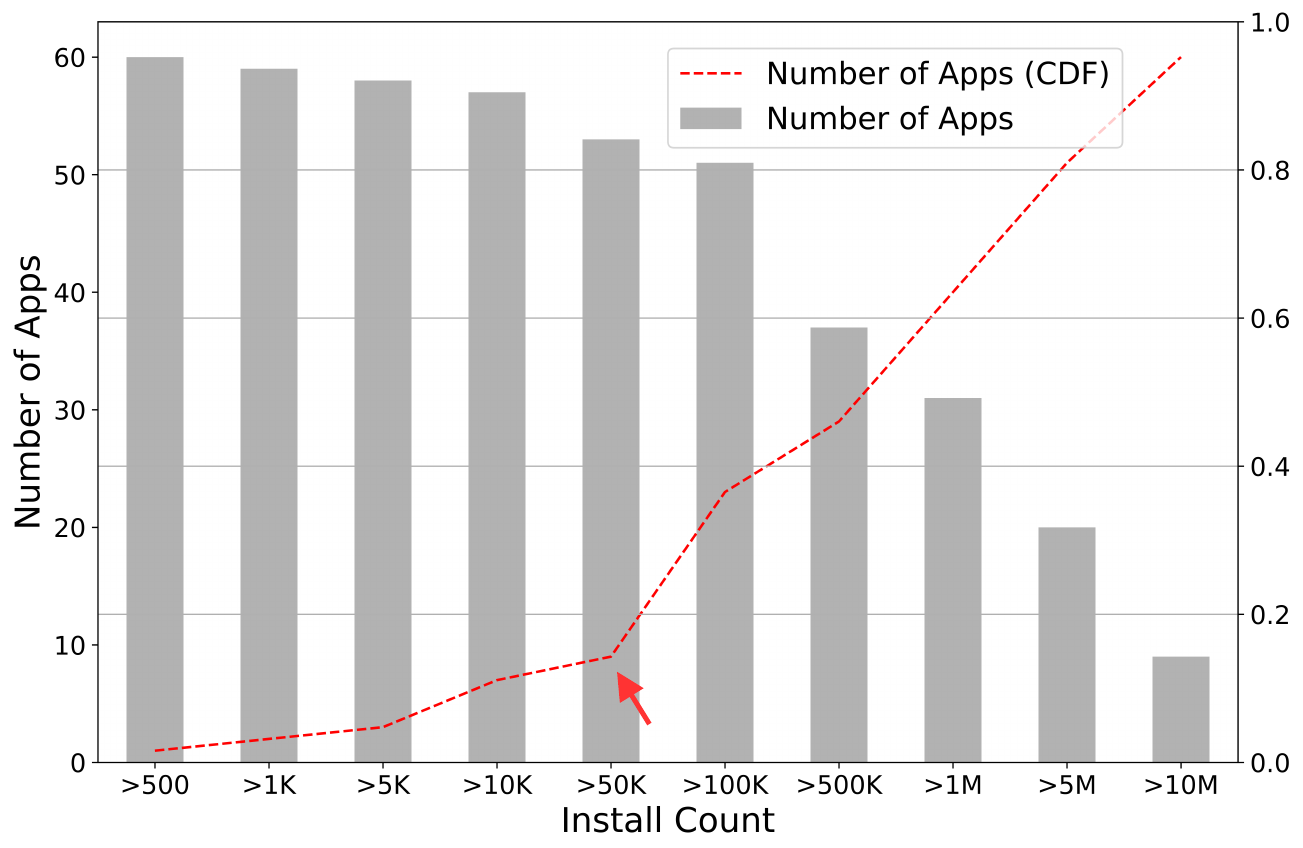}
    \caption{\textbf{Distribution and CDF plot of install count for the 60 shortlisted install-incentivizing apps that collectively account for over 160.5M installs. Eighty-five percent of these apps have 100K or more installs, demonstrating their popularity.}}
    \label{fig:cdf_installs}
\end{minipage}
\hfill
\begin{minipage}{.48\textwidth}
  \centering
    \includegraphics[width=0.75\textwidth]{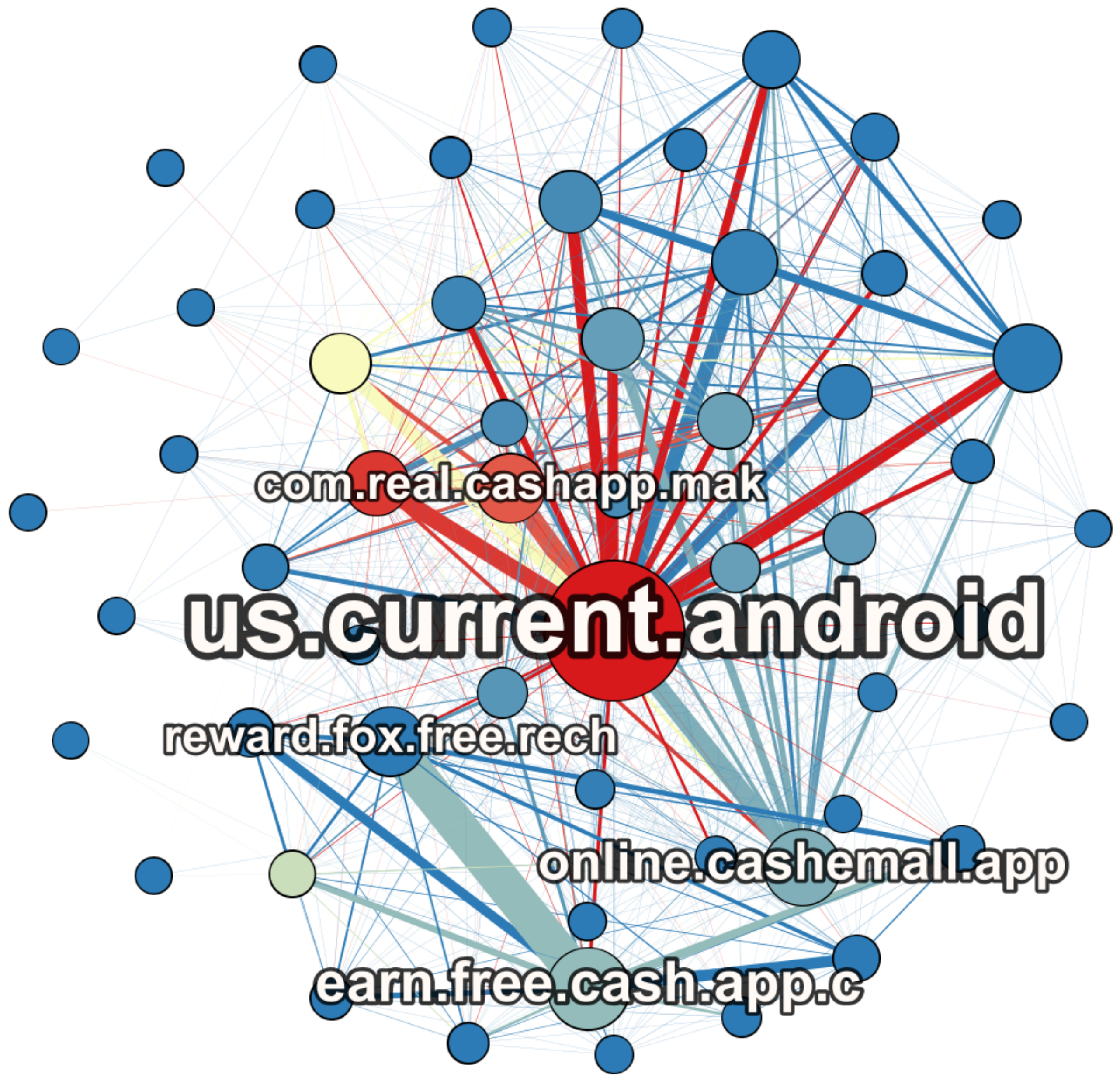}
    \caption{\textbf{Network of apps showing labels of five apps that share the most reviewers with other apps. App `us.current.android' shares 6.4K reviewers with other install-incentivizing apps.}}
    \label{fig:appnetwork}
\end{minipage}%
\vspace{-0.5mm}
\end{figure}

\section{Dataset}
We created queries by prefixing “install apps” to phrases like “earn money”, “win prizes”, “win rewards”, etc., and searched them on Google Play to curate a list of potentially install-incentivizing apps. Then, we proceeded to install the apps from this list on our mobile devices to manually verify whether these apps incentivized installs for other apps; we discarded the apps that did not fit this criterion. Following this process, we shortlisted 60 \textit{install-incentivizing} apps. In Figure~\ref{fig:cdf_installs}, we plot a distribution and CDF of their installs, finding that most apps (85\%) have more than 100K installs. We used a scraper to collect reviews written daily on these apps, over a period of 5 months from November 1, 2021 to April 8, 2022. Reviews were collected daily to avoid over-sampling of reviews from certain temporal periods over others. This resulted in 319,198 reviews from 301,188 reviewers. Figure~\ref{fig:appnetwork} shows a network of apps where edges denote the number of reviewers shared by any two apps. We observe that certain apps share more reviewers with some apps over others, hinting at the possibility of collusion. Lastly, we also collected the permissions requested by apps on users’ devices.

\section{Qualitative Analysis}

To understand the various ways in which install-incentivizing apps affect their users, we performed qualitative analysis of their reviews. Unless a user expands the list of reviews, Google Play displays only the top four “most relevant” reviews under its apps. Owing to their default visibility, we sampled these reviews for all 60 apps over a one-month period, obtaining 1,825 unique reviews. Then, we adopted an inductive open coding approach to thematically code~\cite{miles1994qualitative} these reviews. In the first iteration, all researchers independently worked on identifying high-level codes for these reviews which were then compared and discussed. During this process, we defined the `completion of offers on install-incentivizing apps' as an act of \textit{labor} by users and the `incentive promised for their labor' as \textit{value}. Then, we reached a consensus on four high-level themes: \textit{exploitation, UI challenges, satisfaction}, and \textit{promotion}, which we define below:

\begin{enumerate}
    \item \textbf{Exploitation:} User invests \textit{labor} but is unable to gain \textit{value}.
    \item \textbf{UI challenges:} User invests \textit{labor} but the app's UI makes it challenging for them to gain \textit{value}.
    \item \textbf{Satisfaction:} User invests \textit{labor} and is able to gain \textit{value}. 
    \item \textbf{Promotion:} User invests \textit{labor} in promoting an app through their review, rating or a referral code to gain \textit{value}. 
\end{enumerate}

While all themes were useful for capturing the inter-relationship between a user's \textit{labor} and its \textit{value}, the first three themes were relatively more prevalent in our data. Next, we performed two iterations of line-by-line coding of reviews within the high-level themes where the researchers identified emerging patterns under each theme until the principle of saturation was established. 

\begin{table*}%[!b]
\footnotesize
  \caption{\textbf{Different types of dark patterns mapped to their individual \{Finanical Loss (\textbf{I1}), Invasion of Privacy (\textbf{I2}), Cognitive Burden (\textbf{I3})\} and collective \{Competition (\textbf{C1}), Price Transparency (\textbf{C2}), Trust in the Market (\textbf{C3})\} normative concerns.}}
  \label{darkpatterns}
\centering
\resizebox{\textwidth}{!}{%
\begin{tabular}{|c|c|l|clllll|}
\hline
\multirow{2}{*}{\textbf{\begin{tabular}[c]{@{}c@{}}High-Level \\ Code\end{tabular}}} &
  \multirow{2}{*}{\textbf{\begin{tabular}[c]{@{}c@{}}Low-Level \\ Code\end{tabular}}} &
  \multicolumn{1}{c|}{\multirow{2}{*}{\textbf{Review}}} &
  \multicolumn{6}{c|}{\textbf{Normative Concerns}} \\ \cline{4-9} 
 &
   &
  \multicolumn{1}{c|}{} &
  \multicolumn{1}{c|}{\textbf{I1}} &
  \multicolumn{1}{c|}{\textbf{I2}} &
  \multicolumn{1}{c|}{\textbf{I3}} &
  \multicolumn{1}{c|}{\textbf{C1}} &
  \multicolumn{1}{c|}{\textbf{C2}} &
  \multicolumn{1}{c|}{\textbf{C3}} \\ \hline
\multirow{6}{*}{Exploitation} &
  Withdrawal Limit &
  {\begin{tabular}[c]{@{}l@{}}\textit{100000 is equal to 10 dollars. Just a big waste of time.} \\ \textit{You can not reach the minimum cashout limit.}\end{tabular}} &
  \multicolumn{1}{c|}{\checkmark} &
  \multicolumn{1}{l|}{} &
  \multicolumn{1}{l|}{} &
  \multicolumn{1}{l|}{\checkmark} &
  \multicolumn{1}{l|}{\checkmark} &
  \checkmark \\ \cline{2-9} 
 &
  Cannot Redeem &
  {\begin{tabular}[c]{@{}l@{}}\textit{Absolute scam. Commit time and even made in app} \\ \textit{purchases to complete tasks ... I have over 89k points} \\ \textit{that it refuses to cash out!} \end{tabular}} &
  \multicolumn{1}{c|}{\checkmark} &
  \multicolumn{1}{l|}{} &
  \multicolumn{1}{l|}{} &
  \multicolumn{1}{l|}{\checkmark} &
  \multicolumn{1}{l|}{\checkmark} &
  \checkmark \\ \cline{2-9} 
 &
  Only Initial Payouts &
  {\begin{tabular}[c]{@{}l@{}}\textit{Good for the first one week then it will take forever to} \\ \textit{earn just a dollar. So now I quit this app ...}\end{tabular}} &
  \multicolumn{1}{c|}{\checkmark} &
  \multicolumn{1}{l|}{} &
  \multicolumn{1}{l|}{} &
  \multicolumn{1}{l|}{\checkmark} &
  \multicolumn{1}{l|}{\checkmark} &
  \checkmark \\ \cline{2-9} 
 &
  Paid Offers &
  {\begin{tabular}[c]{@{}l@{}}\textit{In the task I had to deposit 50 INR in an app and I} \\ \textit{would receive 150 INR as a reward in 24 hrs. 5 days}\\ \textit{have passed and I get no reply to mail.}\end{tabular}} &
  \multicolumn{1}{c|}{\checkmark} &
  \multicolumn{1}{l|}{} &
  \multicolumn{1}{l|}{} &
  \multicolumn{1}{l|}{\checkmark} &
  \multicolumn{1}{l|}{\checkmark} &
  \checkmark \\ \cline{2-9} 
 &
  Hidden Costs &
  {\begin{tabular}[c]{@{}l@{}}\textit{Most surveys say that the user isn’t eligible for them,} \\ \textit{after you complete them! Keep in mind you may not} \\ \textit{be eligible for 90\% of the surveys.}\end{tabular}} &
  \multicolumn{1}{c|}{\checkmark} &
  \multicolumn{1}{l|}{} &
  \multicolumn{1}{l|}{} &
  \multicolumn{1}{l|}{\checkmark} &
  \multicolumn{1}{l|}{\checkmark} &
  \checkmark \\ \cline{2-9} 
 &
  Privacy Violations &
  {\begin{tabular}[c]{@{}l@{}}\textit{Enter your phone number into this app and you’ll be} \\ \textit{FLOODED with spam texts and scams. I might have} \\ \textit{to change my phone number because I unwittingly ...}\end{tabular}} &
  \multicolumn{1}{c|}{} &
  \multicolumn{1}{l|}{\checkmark} &
  \multicolumn{1}{l|}{} &
  \multicolumn{1}{l|}{} &
  \multicolumn{1}{l|}{} &
  \checkmark \\ \hline
\multirow{3}{*}{UI Challenges} &
  Too Many Ads &
  {\begin{tabular}[c]{@{}l@{}}\textit{Pathetic with the dam ads! Nothing but ads!!! Money} \\ \textit{is coming but only pocket change. It’ll be 2022 before} \\ \textit{i reach \$50 to cashout, if then.}\end{tabular}} &
  \multicolumn{1}{c|}{} &
  \multicolumn{1}{l|}{} &
  \multicolumn{1}{l|}{\checkmark} &
  \multicolumn{1}{l|}{\checkmark} &
  \multicolumn{1}{l|}{} &
   \\ \cline{2-9} 
 &
  Progress Manipulation &
  {\begin{tabular}[c]{@{}l@{}}\textit{I redownload the app since the app would crash all the} \\ \textit{time ... I logged in and guess what?? ALL MY POINTS} \\ \textit{ARE GONE.. 12k points all  gone...}\end{tabular}} &
  \multicolumn{1}{c|}{\checkmark} &
  \multicolumn{1}{l|}{} &
  \multicolumn{1}{l|}{\checkmark} &
  \multicolumn{1}{l|}{} &
  \multicolumn{1}{l|}{\checkmark} &
  \checkmark \\ \cline{2-9} 
 &
  Permission Override &
  {\begin{tabular}[c]{@{}l@{}}\textit{When you give it permission to go over other apps it} \\ {actually blocks everything else on your phone from} \\ \textit{working correctly including Google to leave this review.}\end{tabular}} &
  \multicolumn{1}{c|}{} &
  \multicolumn{1}{l|}{} &
  \multicolumn{1}{l|}{\checkmark} &
  \multicolumn{1}{l|}{\checkmark} &
  \multicolumn{1}{l|}{} &
  \checkmark \\ \hline
\end{tabular}}
\end{table*}

\subsection{How Install-Incentivizing Apps affect Users}
In this section, we describe our findings from the qualitative analysis to shed light on how install-incentivizing apps affect their users. More specifically, we elaborate on the commonalities and differences of patterns within high-level codes that we discovered using line-by-line coding to depict how labor invested by users in these apps is not only exploited but also leads to negative consequences for them as well as the platform. 
% \vspace{-1.5mm}

\subsubsection{Dark Patterns}\label{dp}\hfill\\ Dark patterns can be defined as tricks embedded in apps that make users perform unintended actions~\cite{brignull2020types}. We find comprehensive descriptions of dark patterns present within install-incentivizing apps in reviews coded as `exploitation' and `UI challenges'. These patterns make it difficult for users to redeem value for their labor. First, our low-level codes uncover the different types of dark patterns present in reviews of install-incentivizing apps. Then, we ground these types in prior literature~\cite{Mathur2021dark} by utilizing lenses of both individual and collective welfare to highlight their normative concerns. The individual lens focuses on dark patterns that allow developers to benefit at the expense of users whereas the collective lens looks at users as a collective entity while examining expenses. In our case, the former comprises three normative concerns. First, patterns that enable developers to extract labor from users without compensating cause \textbf{financial loss (I1)} to users. Second, cases where the data of users is shared with third parties without prior consent, leading to \textbf{invasion of privacy (I2)}. Third, when the information architecture of apps manipulates users into making certain choices due to the induced \textbf{cognitive burden (I3)}. The lens of collective welfare facilitates understanding of the bigger picture of install-incentivizing apps on Google Play by listing three additional concerns. Due to high \textbf{competition (C1)}, some developers incorporate dark patterns in apps that empower them to `extract wealth and build market power at the expense of users’~\cite{day2020dark} on the platform. In conjunction with their concerns at the individual level, they also pose a serious threat to the \textbf{price transparency (C2)} and \textbf{trust in the market (C3)} of Google Play. In Table~\ref{darkpatterns}, we show these different types of dark patterns mapped to their individual and collective normative concerns using sample reviews from our data.
% \vspace{-1.5mm}

\subsubsection{Evidence of Fraudulent Reviews and Ratings}\label{evidence}\hfill\\ During qualitative analysis, we found that most reviews coded as `satisfaction' were relatively shorter and lacked sufficient context to explain how the app benefitted the user, for e.g. \textit{``Good app”, ``Nice App”, ``Very easy to buy money.”, ``Nice app for earning voucher”}. We performed welch’s \textit{t}-test to validate that the number of words in reviews coded as satisfaction were very highly significantly lower than reviews coded as exploitation or UI challenges ($p<0.001,t=-11.41$). The shorter length of reviews, along with the excessive use of adjectives and unrelatedness to the apps represented key spam-detection signals~\cite{shojaee2015framework}, raising suspicions about their fraudulence. We discovered evidence of the same in reviews coded as `promotion’ -- \textit{``Gets high rating because it rewards people to rate it so’’, ``I rated it 5 stars to get credits”}, thus finding that install-incentivizing apps also violate Google Play’s policy by incentivizing users to boost their ratings and reviews. Other reviews coded as `promotion’ involved users promoting other competitor apps (\textit{``No earning 1 task complete not give my wallet not good ! CASHADDA App is good fast earning is good go install now thanks”}) or posting their referral codes to get more credits within the install-incentivizing app (\textit{`The app is Awesome. Use My Referral Code am****02 to get extra coin`”}).

\section{Quantitative Analysis}

In this section, we ascertain findings from our qualitative analysis as well as reveal more characteristics about the behavior of install-incentivizing apps and their reviews. For the same, we examine the permissions requested by these apps to establish their relevance to the dark patterns discussed in Section~\ref{dp}, and perform anomaly detection on their reviews to build upon the evidence of fraud from Section~\ref{evidence}.

\subsection{Permissions in Install-Incentivizing Apps}

\begin{figure}[!h]
\centering
\begin{minipage}{.49\textwidth}
  \centering
    \includegraphics[width=\textwidth]{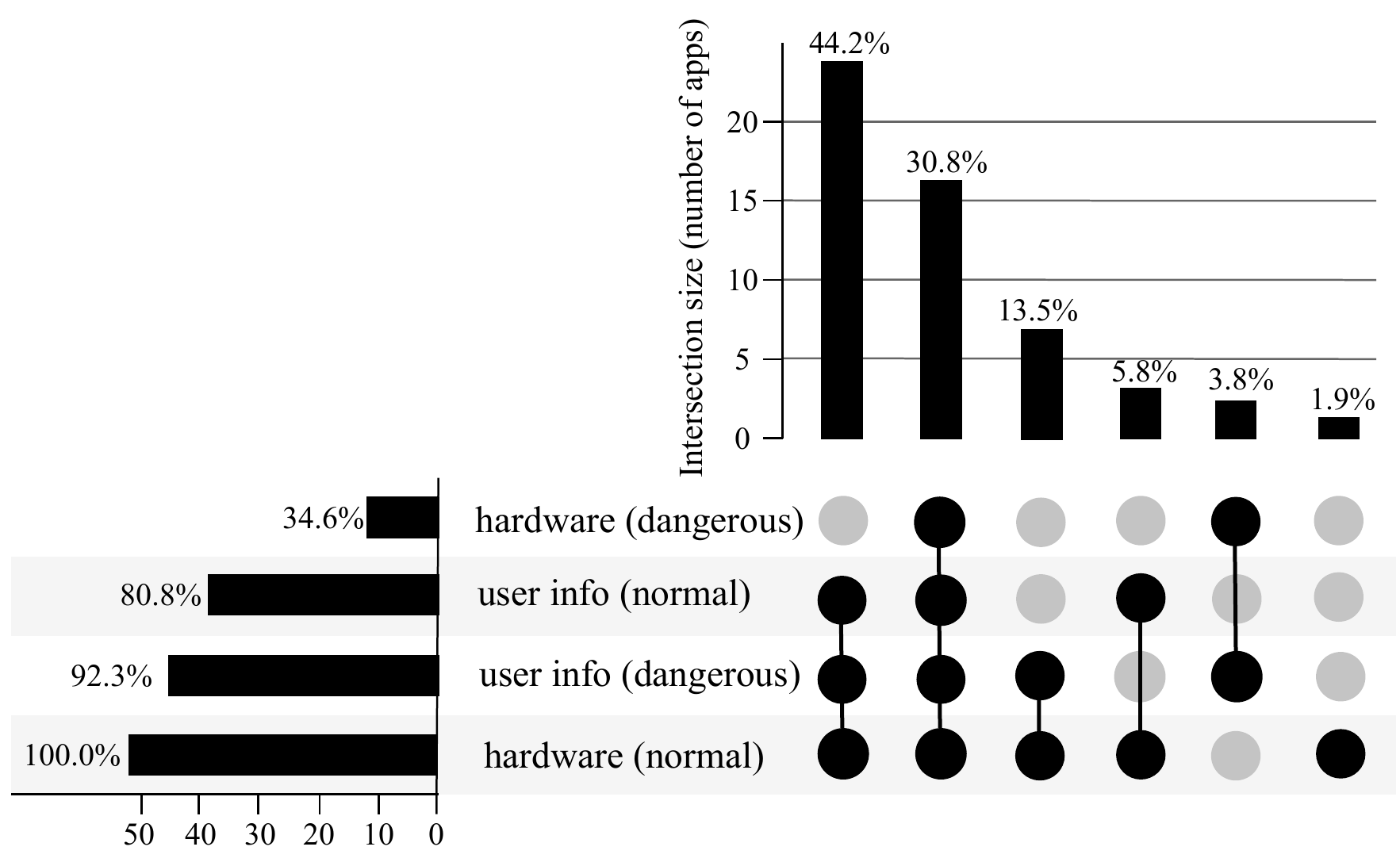}
    \caption{\textbf{UpSet plot demonstrating different types of permissions present in install-incentivizing apps. Over ninety two percent of apps request permissions that access sensitive user information.}}
    \label{fig:upset}
\end{minipage}
\hfill
\begin{minipage}{.49\textwidth}
  \centering
    \includegraphics[width=0.8\textwidth]{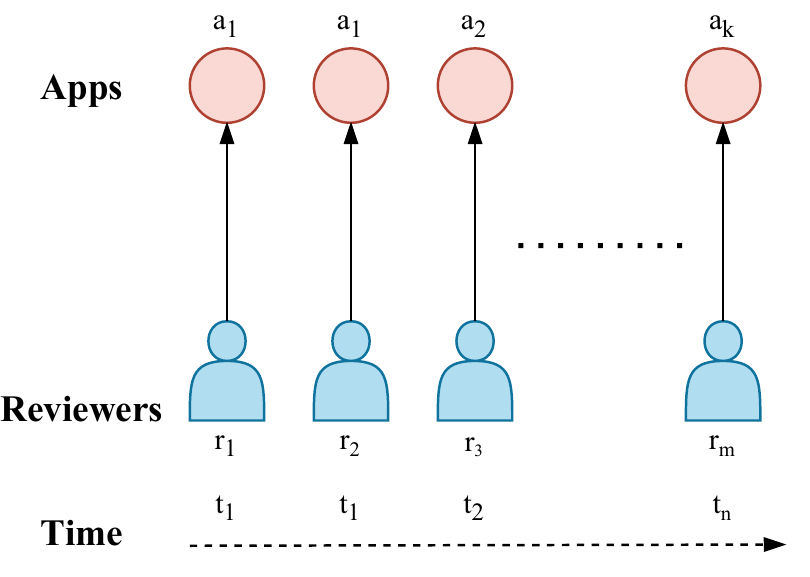}
    \caption{\textbf{Reviews are modelled as an edge-stream in a dynamic bipartite graph of apps and reviewers. Each edge $e \in E$ represents a tuple $(r,a,t)$ where $r$ is a reviewer who reviews an app $a$ at time $t$.}}
    \label{fig:edgestream}
\end{minipage}%
\end{figure}

App permissions support user privacy by protecting access to restricted data and restricted actions on a user’s device \cite{android}.
%~\footnote{\url{https://developer.android.com/guide/topics/permissions/overview}}
Most permissions fall into two protection levels as determined by Android, namely \textit{normal} and \textit{dangerous}, based on the risk posed to user privacy.  Similarly, another distinction can be made between permissions that access \textit{user information} and permissions that only \textit{control device hardware} \cite{pew}. We leverage these categories in our analysis to identify types of permissions prominent across install-incentivizing apps. Figure~\ref{fig:upset} shows an UpSet plot~\cite{2014_infovis_upset} of different types of permissions present in install-incentivizing apps. First, we observe that over 92\% of apps comprise \textit{dangerous} permissions that access user information. The most popular permissions in this category include `modify or delete the contents of your USB storage’ (41 apps), `read phone status and identity’ (24 apps), `access precise location’ (19 apps) and `take pictures and videos’ (14 apps).  Second, despite being requested by relatively fewer apps, some permissions in this category enable an alarming degree of control over user information; for e.g. `create accounts and set passwords’ (5 apps), `add or modify calendar events and send email to guests without owners' knowledge’ (3 apps) and `read your contacts’ (2 apps). Third, 34\% of install-incentivizing apps contain permissions that access dangerous hardware-level information, the most prominent one being `draw over other apps’ (14 apps). Fourth, we note that all but three apps request at least one dangerous permission. Lastly, permissions requested by install-incentivizing apps share common characteristics with the dark patterns discussed above, thus validating their qualitative discovery. 

\subsection{Lockstep Behaviors}

In Section~\ref{evidence}, we found evidence of install-incentivizing apps indulging in review and rating fraud. Thus, we build upon the same to investigate reviews of these apps for anomalous behaviors such as lockstep that are indicative of fraud. Specifically, we focus on detecting groups of reviews that exhibit similar temporal and rating patterns; for e.g. bursts of reviews on an app within a short period of time to boost its overall rating. 

\subsubsection{Modelling and Experimental Setup}\hfill\\
Given that reviews are a temporal phenomenon, we model them as an edge-stream $E = \{e_{1},e_{2},...\}$ of a dynamic graph $G$. Each edge $e_{i} \in E$ represents a tuple $(r_{i},a_{i},t_{i})$ where $r_{i}$ is a reviewer who reviews an app $a_{i}$ at time $t_{i}$ (see Fig~\ref{fig:edgestream}). Groups of fraudulent reviewers may either aim to boost the overall rating of an install-incentivizing app or sink the rating of a competitor app. Thus, we partition our edge stream into two sub-streams as follows: 

\begin{enumerate}
    \item \textbf{$E_{boost}$} $= \{(r_{i},a_{i},t_{i}) \in E\:| \text{ Score}(r_{i},a_{i}) \geq R_{a_{i}} \} \text{, } |E_{boost}|=215,759$ 
    \item \textbf{$E_{sink}$} $= \{(r_{i},a_{i},t_{i}) \in E\:| \text{ Score}(r_{i},a_{i}) < R_{a_{i}} \} \text{, } |E_{sink}|=103,439$ 
\end{enumerate}

where $\text{Score}(r_{i},a_{i}) \in \{1,2,3,4,5\}$ is the score assigned by reviewer $r_{i}$ to the app $a_{i}$ and $R_{a_{i}}$ denotes the overall rating of app $a_{i}$. Next, we reconfigure a state-of-the-art microcluster anomaly detection algorithm \textsc{Midas-F}~\cite{bhatia2020realtime} for our use. In particular, we modify the definition of a microcluster to accommodate the bipartite nature of our dynamic graph. Given an edge $e \in E$, a detection period $T \geq 1$ and a threshold $\beta > 1$, there exists a microcluster of reviews on an app $a$ if it satisfies the following equation:

\begin{equation}
    \begin{split}
    \MoveEqLeft
        \frac{c(e,(n+1)T)}{c(e,nT)} > \beta \text{ where }c(e,nT) = \\ 
        & \bigl\lvert\{(r_{i},a,t_{i}) \mid (r_{i},a,t_{i}) \in E_{boost}
    \land (n-1)T < t_{i} \leq nT\}\bigl\lvert
    \end{split}
\end{equation}

if $e \in E_{boost}$ and vice versa for $E_{sink}$. Depending on whether $e$ is a boosting or sinking edge, $c(e,nT)$ counts similar edges for the app $a$ within consecutive detection periods $(n-1)T$ and $nT$. Values recommended by the authors are used for the remaining parameters $\alpha$ and $ \theta$. It is worth noting that our modification preserves its properties of (i) theoretical guarantees on false positive probability, and (ii) constant-time and constant-memory processing of new edges~\cite{bhatia2020realtime}.

\subsubsection{Analysis and Preliminary Results}
\begin{figure}%[!t]
%\centering
%\begin{minipage}{.49\textwidth}
%  \centering
%    \centering
%    \captionsetup{width=\textwidth}
    \includegraphics[width=0.45\textwidth]{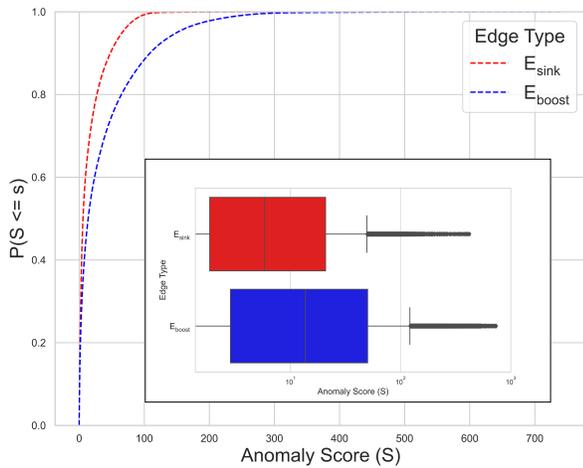}
    \caption{\textbf{CDF plot of anomaly scores for the two edge streams $E_{boost}$ and $E_{sink}$. Reviews that boost the overall rating of an install incentivizing app exhibit significantly more anomalous behavior than reviews that aim to bring it down. }}
    \label{fig:cdfbox}
\end{figure}

\hfill\\
\textsc{Midas-F} follows a streaming hypothesis testing approach that determines whether the observed and expected mean number of edges for a node at a given timestep are significantly different. Based on a chi-squared goodness-of-fit test, the algorithm provides anomaly scores $\text{S}(e)$ for each edge $e$ in a streaming setting. Upon computing anomaly scores for both sub-streams $E_{boost}$ and $E_{sink}$, we visualize their CDF with an inset box plot in Fig~\ref{fig:cdfbox}. It can be observed that $E_{boost}$ exhibits more anomalous behavior than $E_{sink}$. To ascertain statistical significance of the same, we make use of Welch's t-test for the hypothesis $H_{1}: \text{S}_{\mu}(E_{boost}) > \text{S}_{\mu}(E_{sink})$. We infer that reviews that aim to boost the rating of an install-incentivizing app show anomalous behavior that is highly significantly more ($t = 157.23, p<0.0$) than reviews that aim to bring it down. 

\begin{figure}%{.49\textwidth}
%  \centering
%  \captionsetup{width=\textwidth}
    \includegraphics[width=0.44\textwidth]{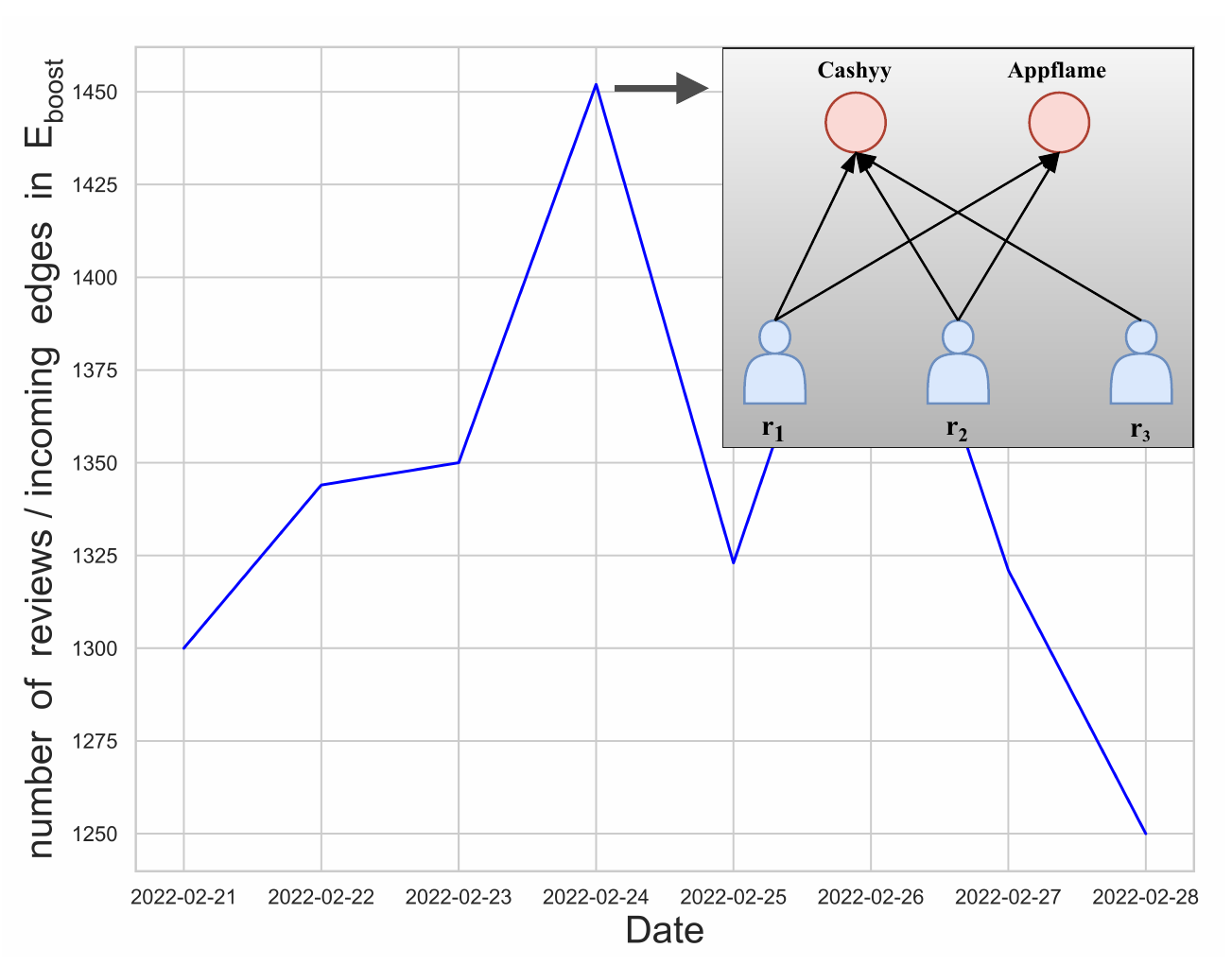}
    \caption{\textbf{A microcluster anomaly detected by the algorithm where three reviewers are boosting the overall rating of two install-incentivizing apps `Cashyy' and `Appflame' on the same day.}}
    \label{fig:toy}
\end{figure}

Next, we examine fraud across anomalous microclusters detected by the algorithm. Figure~\ref{fig:toy} shows one such microcluster anomaly where the algorithm detects reviews from three reviewers boosting the overall rating of two install-incentivizing apps on the same day. We extract the 50 most suspicious clusters of reviews from both sub-streams $E_{boost}$ and $E_{sink}$ based on their average anomaly scores. For each pair of reviews $(r_{i},r_{j})$ within these clusters, we compute their cosine similarity $CS(r_{i},r_{j})$ using embeddings generated by Sentence-BERT~\cite{reimers-2019-sentence-bert}. Over 35\% of reviews (1,687 of 4,717) from the suspicious clusters in $E_{boost}$ form at least one pair of highly identical reviews i.e., $CS(r_{i},r_{j}) = 1$. However, this percentage drops to 10\% (45 of 432 reviews) in case of $E_{sink}$. On closer inspection, we find that these are all extremely short reviews with at most three to four words that comprise mostly of adjectives; for e.g., $E_{boost}$: (`good app', `very good app'), (`good earning app', `very good for earning app'), (`best app', `very best app') and $E_{sink}$: (`bad', `very bad'), (`super', `super'), (`nice', `very nice'). It is surprising to see that all but four identical pairs from $E_{sink}$ contain only positive adjectives considering they assign the app a low rating. A potential reason for this dissonance can be that reviewers writing these reviews want to camouflage as normal users in terms of their rating patterns. Lastly, from the fifty most suspicious clusters, we find such pairs across 47 (94\%) clusters from $E_{boost}$ and 21 (42\%) clusters from $E_{sink}$. This demonstrates that the efficacy of our approach towards detecting lockstep behaviors is not only limited to the temporal and rating dimensions, but also extends to the content present in reviews.

%\hfill

\section{Discussion and Future Work}

Our current work sheds light on how lax implementation of Google Play’s policy on fraudulent installs, ratings and reviews empowers developers of install-incentivizing apps to deplete the trust and transparency of the platform. Through use of permissions that access restricted data and perform restricted actions, developers incorporate dark patterns in these apps to deceive users and extort labor from them in the form of offers. The second form of labor that we study in our work is the writing of fraudulent reviews. We find evidence of their presence qualitatively and show promising results in detecting them algorithmically. Both types of fraud (incentivized installs and reviews) are only made possible by the labor of users who are vulnerable or crowd-workers who are underpaid~\cite{aso2019}. This enables developers to extract profits as they get away with violating Google Play’s policies without any consequences or accountability. However, a question that remains unanswered is, if reviews under these apps describe exploitative experiences of users, what is it that facilitates their continued exploitation? For now, we can only conjecture that fraudulent positive reviews on install-incentivizing apps suppress ranks of reviews containing exploitative experiences of users. Whether the same holds true or not is a question that remains to be explored in our future work.

% \vspace{-1.0mm}

\bibliographystyle{ACM-Reference-Format}
\bibliography{sample-base}

\end{document}